\documentclass[10pt,conference,english,twocolumn]{IEEEtran}
\IEEEoverridecommandlockouts

\usepackage{babel}
\usepackage[babel]{microtype}
\usepackage[babel]{csquotes}

\usepackage[svgnames]{xcolor}
\usepackage{graphicx}

\usepackage{amsmath}
\usepackage{amsfonts}
\usepackage{amssymb}
\usepackage{amsthm}
\usepackage{algorithm,algpseudocode}
\usepackage{bm}
\usepackage{siunitx}
\usepackage{booktabs}
\usepackage{subfigure}
\usepackage{geometry}
\usepackage{tikz}
\usepackage{pgfplots}
\pgfplotsset{compat=newest}
\usetikzlibrary{positioning}
\usetikzlibrary{shapes.geometric, arrows, decorations.pathreplacing}
\usetikzlibrary{shapes.arrows}
\usetikzlibrary{3d}

\usepackage[defernumbers=true,
backend=biber,
isbn=false,
url=false,
maxbibnames=999,
maxcitenames=10,
doi=false,
giveninits=true,
sortcites=true,
eprint=true,
hyperref=true,
style=ieee,
bibstyle=ieee-proc,
]{biblatex}
\bibliography{modernize-ref.bib}

\usepackage{hyperref}
\hypersetup{%
  pdftitle = {Distributed Combinatorial Optimization of User Assignment in mmWave Cell-free Massive MIMO with Graph Neural Network},
  pdfauthor = {Bile Peng, Bihan Guo, Karl-Ludwig Besser, Luca Kunz, Ramprasad Raghunath, Anke Schmeink, Eduard A. Jorswieck, Giuseppe Caire, H. Vincent Poor},
  pdflang = {en-US},
  colorlinks,
  allcolors=.,
  urlcolor=MidnightBlue,
}

\usepackage[acronyms,nomain,xindy,nowarn]{glossaries}
\makeglossaries
\newacronym{isac}{ISAC}{integrated sensing and communication}
\newacronym{ae}{AE}{auto-encoder}
\newacronym{vae}{VAE}{variational auto-encoder}
\newacronym{ap}{AP}{access point}
\newacronym[plural=AoA, firstplural=angles-of-arrival (AoA)]{aoa}{AoA}{angle-of-arrival}
\newacronym{ao}{AO}{alternating optimization}
\newacronym{bs}{BS}{base station}
\newacronym{dnn}{DNN}{deep neural network}
\newacronym{cnn}{CNN}{convolutional neural network}
\newacronym{rnn}{RNN}{recurrent neural network}
\newacronym{ma}{MA}{multiple access}
\newacronym{nn}{NN}{neural network}
\newacronym{ue}{UE}{user equipment}
\newacronym{noma}{NOMA}{non-orthogonal multiple access}
\newacronym{mimo}{MIMO}{multiple-input-multiple-output}
\newacronym{miso}{MISO}{multiple-input-single-output}
\newacronym{cf}{CF mMIMO}{cell-free massive MIMO}
\newacronym{cdf}{CDF}{cumulative distribution function}
\newacronym{gnn}{GNN}{graph neural network}
\newacronym{cc}{CC}{channel charting}
\newacronym{los}{LoS}{line-of-sight}
\newacronym{csi}{CSI}{channel state information}
\newacronym{star}{STAR}{simultaneously transmitting and reflecting}
\newacronym{gsd}{GSD}{generalized serial dictatorship}
\newacronym{aod}{AoD}{angle-of-departure}
\newacronym{lwf}{LwF}{learing without forgetting}
\newacronym{sinr}{SINR}{signal-to-interference-plus-noise ratio}
\newacronym{snr}{SNR}{signal-to-noise ratio}
\newacronym{sgd}{SGD}{stochastic gradient descent}
\newacronym{agi}{AGI}{artificial general intelligence}
\newacronym{llm}{LLM}{large language model}
\newacronym{dl}{DL}{deep learning}
\newacronym{rl}{RL}{reinforcement learning}
\newacronym{crb}{CRB}{Cram\'er-Rao bound}
\newacronym{tpu}{TPU}{tensor processing unit}
\newacronym{evt}{EVT}{extreme value theory}
\newacronym{urllc}{URLLC}{ultra-reliable low latency communications}
\newacronym{mrt}{MRT}{maximum ratio transmission}
\newacronym{embb}{eMBB}{enhanced mobile broadband}
\newacronym{mmse}{MMSE}{minimum mean square error}
\newacronym{qd}{QD}{quasi degradation}
\newacronym{rss}{RSS}{received signal strength}
\newacronym{siso}{SISO}{single-input single-output}
\newacronym{rsma}{RSMA}{rate-splitting multiple access}
\newacronym{lstm}{LSTM}{long short-term memory}
\newacronym{drl}{DRL}{deep reinforcement learning}
\newacronym{dqn}{DQN}{deep Q-network}
\newacronym{ml}{ML}{machine learning}
\newacronym{sca}{SCA}{successive convex approximation}
\newacronym{admm}{ADMM}{alternating direction method of multipliers}
\newacronym{alm}{ALM}{augmented Lagrangian method}
\newacronym{mmwave}{mmWave}{millimeter wave}
\newacronym{mm}{MM}{majorization-minimization}
\newacronym{rmcg}{RMCG}{Riemannian manifold conjugate gradient}
\newacronym{hpe}{HPE}{hierarchically permutation-equivariant}
\newacronym{cb}{CB}{conjugate beamforming}
\newacronym{sdma}{SDMA}{space-division multiple access}
\newacronym{pa}{PA}{power amplifier}
\newacronym{sdr}{SDR}{semidefinite relaxation}
\newacronym{mse}{MSE}{mean squared error}
\newacronym{ris}{RIS}{reconfigurable intelligent surface}
\newacronym{tof}{ToF}{time of flight}
\newacronym{em}{EM}{expectation-maximization}
\newacronym{pe}{PE}{permutation-equivariance}
\newacronym{zf}{ZF}{zero-forcing}
\newacronym{ekf}{EKF}{extended Kalman filter}
\newacronym{bcd}{BCD}{block coordinate descent}
\newacronym{concrete}{CONCRETE}{CONtinuous relaxation for disCRETE}
\newacronym{mop}{MOP}{multi-objective programming}
\newacronym{uwb}{UWB}{ultra-wide band}
\newacronym{sic}{SIC}{successive interference cancellation}
\newacronym{otfs}{OTFS}{orthogonal time frequency space}
\newacronym{ofdm}{OFDM}{orthogonal frequency-division multiplexing}
\newacronym{mpc}{MPC}{multi-path component}
\newacronym{wmmse}{WMMSE}{weighted minimum mean square error}
\newacronym{wsr}{WSR}{weighted sum-rate}
\newacronym{mab}{MAB}{multi-armed bandit}
\newacronym{ppo}{PPO}{proximal policy optimization}
\newacronym{sac}{SAC}{soft actor-critic}
\newacronym{iid}{i.i.d.}{independent and identically distributed}
\newacronym{wrt}{w.r.t.}{with respect to}
\setacronymstyle{long-short}
\glsdisablehyper

\addto\extrasenglish{%

}

\newcommand{\todo}[2][]{\ignorespaces
	\if\relax\detokenize{#1}\relax
	{\color{red}[TODO: #2]}%
	\else
	{\color{red}[TODO (#1): #2]}%
	\fi
}

\definecolor{plot0}{HTML}{004488}
\definecolor{plot1}{HTML}{DDAA33}
\definecolor{plot2}{HTML}{BB5566}
\definecolor{plot3}{HTML}{000000}
\definecolor{plot4}{HTML}{AAAAAA}
\definecolor{plot0comp}{HTML}{C17A00}

\usepackage{xspace}

\hypersetup{draft}

\title{Distributed Combinatorial Optimization of Downlink User Assignment in mmWave Cell-free Massive MIMO Using Graph Neural Networks}
\author{
\IEEEauthorblockN{%
Bile Peng\IEEEauthorrefmark{1}, 
Bihan Guo\IEEEauthorrefmark{1}, 
Karl-Ludwig Besser\,\IEEEauthorrefmark{3}, 
Luca Kunz\IEEEauthorrefmark{1},
Ramprasad Raghunath\IEEEauthorrefmark{1}, 
Anke Schmeink\IEEEauthorrefmark{2},\\
Eduard A. Jorswieck\IEEEauthorrefmark{1},
Giuseppe Caire\IEEEauthorrefmark{4},
H. Vincent Poor\,\IEEEauthorrefmark{3}}

\IEEEauthorblockA{\IEEEauthorrefmark{1}Institute for Communications Technology, Technische Universit\"at Braunschweig, Germany}

\IEEEauthorblockA{\IEEEauthorrefmark{3}Department of Electrical and Computer Engineering, Princeton University, USA}

\IEEEauthorblockA{\IEEEauthorrefmark{2}Chair of Information Theory and Data Analytics, RWTH Aachen University, Germany}

\IEEEauthorblockA{\IEEEauthorrefmark{4}Communications and Information Theory Group, Technische Universit\"at Berlin, Germany}

Email: \{b.peng, bihan.guo, luca.kunz, r.raghunath, e.jorswieck\}@tu-braunschweig.de, \{karl.besser, poor\}@princeton.edu,\\anke.schmeink@inda.rwth-aachen.de, caire@tu-berlin.de
\thanks{The work of B.~Peng, R.~Raghunath, A.~Schmeink, E.~Jorswieck and G.~Caire is supported by the Federal Ministry of Education and Research Germany (BMBF) as part of the 6G Research and Innovation Cluster (6G-RIC) under Grant 16KISK031.
The work of K.-L.~Besser is supported by the German Research Foundation (DFG) under grant BE\,8098/1-1. The work of L.~Kunz is supported by the BMBF under grant 16KISK074.
The work of H.~V.~Poor was supported in part by the U.S National Science Foundation under Grants CNS-2128448 and ECCS-2335876.}
}

\begin{document}

\maketitle

\begin{abstract}
\Gls{mmwave} \gls{cf} is a promising solution for future wireless communications.
However, its optimization is non-trivial due to the challenging channel characteristics.
We show that \gls{mmwave} \gls{cf} optimization is largely an assignment problem between \glspl{ap} and users
due to the high path loss of \gls{mmwave} channels,
the limited output power of the amplifier,
and the almost orthogonal channels between users
given a large number of \gls{ap} antennas.
The combinatorial nature of the assignment problem,
the requirement for scalability,
and the distributed implementation of \gls{cf} make this problem difficult.
In this work,
we propose an unsupervised \gls{ml} enabled solution.
In particular, a \gls{gnn} customized for scalability and distributed implementation is introduced.
Moreover, the customized \gls{gnn} architecture is \gls{hpe},
i.e., if the \glspl{ap} or users of an \gls{ap} are permuted, the output assignment is automatically permuted in the same way.
To address the combinatorial problem,
we relax it to a continuous problem, and introduce an information entropy-inspired penalty term.
The training objective is then formulated using the \gls{alm}.
The test results show that the realized sum-rate outperforms that of the \gls{gsd} algorithm
and is very close to the upper bound in a small network scenario,
while the upper bound is impossible to obtain in a large network scenario.
\end{abstract}

\begin{IEEEkeywords}
Assignment,
augmented Lagrangian method,
cell-free massive MIMO,
graph neural network,
unsupervised machine learning.
\end{IEEEkeywords}

\glsresetall
\section{Introduction}
\label{sec:intro}

\Gls{cf} is a promising solution for next-generation wireless communications.
By deploying \glspl{ap} in a high density and
enabling cooperation between \glspl{ap} to serve the same user,
\gls{cf} can realize high \gls{rss},
low interference thanks to the cooperation between \glspl{ap},
and high reliability due to the resulting macro diversity~\cite{ammar2021user}.
These properties make \gls{cf} suitable for \gls{mmwave} communication,
which enjoys a large bandwidth
but suffers from low \gls{rss} due to high propagation loss and low amplifier output power.
In addition, weak diffraction at high frequencies~\cite{virk2019modeling} makes blockage a serious problem.
By using the \gls{mmwave} spectrum and cooperation of multiple \glspl{ap} in \gls{cf},
the \gls{rss} is enhanced and the blockage problem is mitigated through macro-diversity.

Given the hardware and channel properties, optimizing \gls{mmwave} \gls{cf} is primarily an assignment problem between \glspl{ap} and users.
Sophisticated precoding techniques,
such as \gls{zf} and \gls{mmse},
are neither feasible (due to the immature hardware)
nor indispensable (the spectral efficiency is not crucial given a large bandwidth).
Instead, conjugate precoding 
is more feasible with simple hardware,
realizes high antenna gain given many antennas,
and reduces interference significantly because multi-user channels become asymptotically orthogonal as the number of antennas increases~\cite{marzetta2016fundamentals}.
Moreover, 
transmit power control is not necessary because 
the optimal transmit power is always the maximum available power
given the high path loss
for typical objectives.

In the literature,
a mixed-integer programming problem for \gls{cf} fronthaul resource allocation is considered in~\cite{li2023joint}.
\Glspl{ap} and pilots are assigned in a rule-based, distributed way in~\cite{bjornson2020scalable}.
An alternative selection method is proposed for \gls{ap}-user assignment in~\cite{sarker2023access}.
The block-sparsity norm $l_{2, 1}$ is applied to optimize \gls{ap}-user assignment and power control simultaneously in~\cite{vu2020joint}.
The pilot assignment problem is considered based on serving set in~\cite{zaher2023soft}
and graph coloring in~\cite{liu2020graph}.

It can be concluded from the above that the assignment problem in \gls{cf} optimization 
is difficult to solve due to its non-differentiability.
The existing algorithms are either heuristic or do not scale well.
In recent years,
\gls{ml} has been proposed to solve combinatorial problems in~\cite{bengio2021machine}.
Such methods train a \gls{nn} on massive amounts of data to find an optimized mapping from problem representation to solution.
In the literature,
an encoder-decoder like \gls{nn} is applied to optimize \gls{ap} scheduling in supervised manner in~\cite{guenach2021deep}.
In \cite{khalil2017learning}, a \enquote{meta-algorithm} is proposed to generate a greedy policy for combinatorial problems by joining \gls{rl} and graph embedding.
\Gls{rl} is applied in~\cite{oh2023decentralized} to optimize pilot assignment in a decentralized way.
However, if we do not have \enquote{correct} labels and must use unsupervised learning for optimization,
trial-and-error based \gls{rl} is not necessarily the optimal choice,
since it has a poor scalability due to the \enquote{curse of dimensionality}~\cite{verleysen2005curse}.
On the other hand, if the objective is known, model-based optimization is a better choice because it does not rely on sampling solutions through trial-and-error.
Therefore, the dimensionality has only a moderate impact.

Among different contributions to \gls{cf} optimization with \gls{ml},
the \gls{gnn} architecture has gained particular attention due to its distributed nature and high scalability~\cite{ranasinghe2021graph,li2023heterogeneous,raghunath2024energy,salaun2022gnn}.
Despite its significant potential,
combinatorial assignment problems with connection number constraints remain open to the best of our knowledge.

We propose a model-based \gls{ml} approach to optimize the assignment between \glspl{ap} and users
using a customized \gls{gnn}.
Compared to existing works,
the proposed solution is unsupervised (i.e., it does not need \enquote{correct} labels)
and model-based.
Our main contributions are summarized below.
\begin{itemize}
\item We propose a customized \gls{hpe}~\cite{umagami2023hiperformer} \gls{gnn},
i.e., any permutation of \glspl{ap} and users leads to the same permutation of the solution.
Moreover,
data traffic in the fronthaul between \glspl{ap} (i.e., message passing between nodes of the \gls{gnn})
is reduced significantly compared to the canonical \gls{gnn}.
\item We relax the original combinatorial problem to a continuous one,
design a penalty inspired by information entropy to enforce discreteness,
and apply the \gls{alm} for training.
\end{itemize}
Beyond the \gls{ap}-user assignment problem, 
this work is an early contribution to unsupervised \gls{ml} for general scalable combinatorial optimization problems.
\section{Problem Statement}
\label{sec:problem}

\subsection{Problem Formulation}
\label{sec:formulation}

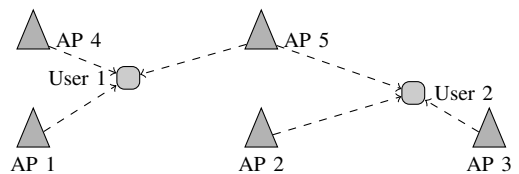
\begin{figure}[b]
    \centering
    \begin{tikzpicture}
    \tikzstyle{every node}=[font=\footnotesize]
    \tikzstyle{ap}=[isosceles triangle, draw, rotate=90, fill=gray!60, minimum size =.25cm]
    \tikzstyle{user}=[rectangle, draw, fill=gray!40, minimum size =.3cm, rounded corners=0.1cm]
	
    \node[ap,label={left:AP 1}] (ap1) at (0,0){};
    \node[ap,label={left:AP 2}] (ap2) at (3,0){};
    \node[ap,label={left:AP 3}] (ap3) at (6,0){};
    \node[ap,label={below:AP 4}] (ap4) at (0,1.3){};
    \node[ap,label={below:AP 5}] (ap5) at (3,1.3){};
 
    \node[user,label={left:User 1}] (u1) at (1.25,0.8){};
    \node[user,label={right:User 2}] (u2) at (5, 0.6){};

    \draw[dashed, ->] (ap1) -- (u1);
    \draw[dashed, ->] (ap4) -- (u1);
    \draw[dashed, ->] (ap5) -- (u1);
    \draw[dashed, ->] (ap2) -- (u2);
    \draw[dashed, ->] (ap5) -- (u2);
    \draw[dashed, ->] (ap3) -- (u2);
\end{tikzpicture}
    \caption{The \gls{mmwave} \gls{cf} network scenario in the downlink, 
    where dashed arrows indicate assignments between \glspl{ap} and users.}
    \label{fig:scenario}
\end{figure}

We consider a \gls{mmwave} \gls{cf} network
with $N$~\glspl{ap} and $K$~users,
as illustrated in \autoref{fig:scenario}.

In this \gls{cf} network,
every \gls{ap} is equipped with a large antenna array with an analog beamformer,
which performs conjugate precoding.
Due to the narrow beamwidth and height difference between \glspl{ap} and users,
interference between users is negligible even with similar azimuths of arrival\footnote{The elevations of arrival are different for users with different distances to their common serving \gls{ap}.}.
Therefore, we make the pragmatic assumption that interference can be omitted.
Moreover,
due to high propagation path loss at \gls{mmwave} frequencies and the limited output power of the amplifier,
we assume that the optimal transmit power is the maximum available power.
This transforms the conventional precoding and transmit power control problem into a combinatorial assignment problem:
The optimization variable~$s_{kn} \in \{0, 1\}$ indicates 
whether \gls{ap}~$n$ serves user~$k$ ($s_{kn} = 1$) or not ($s_{kn}=0$).
Since phase synchronization is difficult for \gls{mmwave} communications, 
we assume non-coherent phases between \glspl{ap}.
The capacity of user~$k$ is therefore
\begin{equation}
    r_k = \log_2\left( 1 + \frac{\sum_{n=1}^N g_{kn}s_{kn}}{\sigma^2} \right),
    \label{eq:rate}
\end{equation}
where $g_{kn}$ is the effective channel gain from \gls{ap}~$n$ to user~$k$ including transmit power, antenna gain and path loss, and $\sigma^2$ is the noise power (we assume the noise is additive white and Gaussian).

We assume that
each \gls{ap} can serve at most $U$~users due to hardware constraints, and
each user must be served by at least $L$~\glspl{ap} to ensure reliability via macro-diversity.
Our objective is to maximize the sum rate.
The problem is formulated as
\begin{subequations}
\begin{align}
    \max_{\mathbf{S}} \quad & f=\sum_{k=1}^K r_k \label{eq:objective}\\
    \text{s.t.}\quad & s_{kn} \in \{0, 1\},\label{eq:discrete} \\
    & \sum_{k=1}^K s_{kn} \leq U, \quad \forall n\label{eq:upper}\\
    & \sum_{n=1}^N s_{kn} \geq L, \quad \forall k\label{eq:lower}
\end{align}
\label{eq:problem}%
\end{subequations}
where $\mathbf{S}\in \{0, 1\}^{K\times N}$ has element~$s_{kn}$ in row~$k$ and column~$n$.

\subsection{Problem Properties}
\label{sec:properties}

We identify two major properties of \eqref{eq:problem} that must be reflected in the proposed solution as follows.

\subsubsection{Hierarchical Permutation-Equivariance}

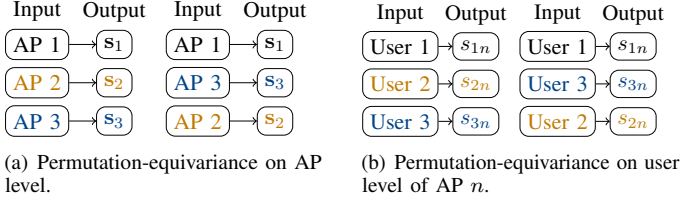
\begin{figure}
    \centering
    \subfigure[Permutation-equivariance on \gls{ap} level.\label{fig:pe_ap}]{\centering\resizebox{.47\linewidth}{!}{\begin{tikzpicture}
\node (1) [rectangle, rounded corners, draw=black, label={above:Input}] at (0, 0) {AP 1};
\node (2) [rectangle, rounded corners, draw=black, below of=1, yshift=.4cm, font=\color{plot0comp}] {AP 2};
\node (3) [rectangle, rounded corners, draw=black, below of=2, yshift=.4cm, font=\color{plot0}] {AP 3};

\node (p1) [rectangle, rounded corners, draw=black, right of=1, xshift=.2cm, label={above:Output}] {$\mathbf{s}_1$};
\node (p2) [rectangle, rounded corners, draw=black, below of=p1, yshift=.4cm, font=\color{plot0comp}] {$\mathbf{s}_2$};
\node (p3) [rectangle, rounded corners, draw=black, below of=p2, yshift=.4cm, font=\color{plot0}] {$\mathbf{s}_3$};

\draw [->] (1) -- (p1);
\draw [->] (2) -- (p2);
\draw [->] (3) -- (p3);

\node (1') [rectangle, rounded corners, draw=black, label={above:Input}] at (2.5, 0) {AP 1};
\node (2') [rectangle, rounded corners, draw=black, below of=1', yshift=.4cm, font=\color{plot0}] {AP 3};
\node (3') [rectangle, rounded corners, draw=black, below of=2', yshift=.4cm, font=\color{plot0comp}] {AP 2};

\node (p1') [rectangle, rounded corners, draw=black, right of=1', xshift=.2cm, label={above:Output}] {$\mathbf{s}_1$};
\node (p2') [rectangle, rounded corners, draw=black, below of=p1', yshift=.4cm, font=\color{plot0}] {$\mathbf{s}_3$};
\node (p3') [rectangle, rounded corners, draw=black, below of=p2', yshift=.4cm, font=\color{plot0comp}] {$\mathbf{s}_2$};

\draw [->] (1') -- (p1');
\draw [->] (2') -- (p2');
\draw [->] (3') -- (p3');

\end{tikzpicture}}}
    \hfill
    \subfigure[Permutation-equivariance on user level of \gls{ap}~$n$.\label{fig:pe_user}]{\centering\resizebox{.47\linewidth}{!}{\begin{tikzpicture}
\node (1) [rectangle, rounded corners, draw=black, label={above:Input}] at (0, 0) {User 1};
\node (2) [rectangle, rounded corners, draw=black, below of=1, yshift=.4cm, font=\color{plot0comp}] {User 2};
\node (3) [rectangle, rounded corners, draw=black, below of=2, yshift=.4cm, font=\color{plot0}] {User 3};

\node (p1) [rectangle, rounded corners, draw=black, right of=1, xshift=.2cm, label={above:Output}] {$s_{1n}$};
\node (p2) [rectangle, rounded corners, draw=black, below of=p1, yshift=.4cm, font=\color{plot0comp}] {$s_{2n}$};
\node (p3) [rectangle, rounded corners, draw=black, below of=p2, yshift=.4cm, font=\color{plot0}] {$s_{3n}$};

\draw [->] (1) -- (p1);
\draw [->] (2) -- (p2);
\draw [->] (3) -- (p3);

\node (1') [rectangle, rounded corners, draw=black, label={above:Input}] at (2.5, 0) {User 1};
\node (2') [rectangle, rounded corners, draw=black, below of=1', yshift=.4cm, font=\color{plot0}] {User 3};
\node (3') [rectangle, rounded corners, draw=black, below of=2', yshift=.4cm, font=\color{plot0comp}] {User 2};

\node (p1') [rectangle, rounded corners, draw=black, right of=1', xshift=.2cm, label={above:Output}] {$s_{1n}$};
\node (p2') [rectangle, rounded corners, draw=black, below of=p1', yshift=.4cm, font=\color{plot0}] {$s_{3n}$};
\node (p3') [rectangle, rounded corners, draw=black, below of=p2', yshift=.4cm, font=\color{plot0comp}] {$s_{2n}$};

\draw [->] (1') -- (p1');
\draw [->] (2') -- (p2');
\draw [->] (3') -- (p3');

\end{tikzpicture}}}
    \caption{The hierarchical permutation-equivariance. If the inputs are permuted, the outputs should be permuted in the same way automatically.}
    \label{fig:pe}
\end{figure}

Problem~\eqref{eq:problem} is \emph{\gls{hpe}}:
On the \gls{ap} level,
if we permute \glspl{ap},
the user assignment of every \gls{ap} should be permuted in the same way,
as illustrated in \autoref{fig:pe_ap},
where $\mathbf{s}_n$ is the $n$-th column of $\mathbf{S}$.
On the user level of every \gls{ap},
if we permute users,
the user assignment of every \gls{ap} should be permuted in the same way,
as illustrated in \autoref{fig:pe_user}.
This property does not exist in conventional \gls{nn} architectures.
In \autoref{sec:gnn},
we propose a \gls{gnn} with specially designed architectures for node and edge processing
such that the \gls{gnn} is \gls{hpe}.

\subsubsection{Non-differentiable Combinatorial Optimization}

Constraint \eqref{eq:discrete} makes \eqref{eq:problem} a combinatorial problem.
With existing methods, 
this problem can be tackled either through exhaustive search for small problem sizes 
or via heuristic approaches with suboptimal performance.
Neither of these approaches is ideal.
Problem~\eqref{eq:problem} is also difficult to solve with \glspl{nn}
because \gls{nn} training is based on gradients.
However, this problem is non-differentiable.
In \autoref{sec:training},
we propose a training process that relaxes \eqref{eq:problem} to a continuous one,
and then uses a penalty and \gls{alm} to find solutions to the original problem.
\section{General Approach of the Unsupervised Machine Learning for Optimization}
\label{sec:framework}

We present the general approach of unsupervised \gls{ml} for optimization as follows.
Given effective channel gains~$\mathbf{G} \in \mathbb{R}^{K\times N}$,
with element~$g_{kn}$ in row~$k$ and column~$n$ as defined in \eqref{eq:rate},
we look for a solution~$\mathbf{S}$ that maximizes objective~$f$ defined in \eqref{eq:objective},
which is fully determined by $\mathbf{G}$ and $\mathbf{S}$, and can be written as
$f(\mathbf{G}, \mathbf{S})$.
We define an \gls{nn}~$N_\theta$, which is parameterized by $\theta$, 
i.e., $\theta$ contains all trainable weights and biases of $N_\theta$.
$N_\theta$
maps from $\mathbf{G}$
to $\mathbf{S}$,
i.e., $\mathbf{S} = N_\theta(\mathbf{G})$.
The objective can then be written as $f(\mathbf{G}, \mathbf{S})=f(\mathbf{G}, N_\theta(\mathbf{G}); \theta)$,
where it is emphasized that $f$ depends on $\theta$.
We collect massive amounts of realizations of $\mathbf{G}$ in a training set~$\mathcal{D}$
and formulate the problem as
\begin{equation}
    \max_\theta J=\sum_{\mathbf{G} \in \mathcal{D}} f(\mathbf{G}, N_\theta(\mathbf{G}); \theta).
    \label{eq:ml}
\end{equation}
In this way, $N_\theta$ is optimized for the ensemble of $\mathbf{G} \in \mathcal{D}$ (\emph{training}) using gradient ascent:
\begin{equation}
    \theta \leftarrow \theta + \eta \nabla_\theta J,
    \label{eq:gradient-ascend}
\end{equation}
where $\eta$ is learning rate.
If $N_\theta$ is well trained,
$\mathbf{S}' = N_\theta (\mathbf{G'})$ is also a good solution for $\mathbf{G}' \notin \mathcal{D}$ (\emph{testing}),
like a human uses experience to solve new problems of the same type\footnote{A complete retraining is only required when the input is fundamentally changed, e.g., when there is a change of deployment environment.}~\cite{yu2022role}.

Compared to analytical methods,
which optimize $\mathbf{S}$ itself given $\mathbf{G}$,
unsupervised \gls{ml} optimizes the mapping from $\mathbf{G}$ to $\mathbf{S}$.
This has the following two advantages:
\begin{enumerate}
    \item \textit{Low complexity in operation:} 
    If analytical optimization is used, it is performed after receiving $\mathbf{G}$.
    If its complexity is high, a real-time application may be difficult.
    In contrast, with \gls{ml}-based optimization, $N_\theta$ is trained prior to operation.
    During operation, using the trained~$N_\theta$ to make inferences on a new $\mathbf{G}'$ is much less complex than training or analytical optimization.
    \item \textit{Scalability:}
    As will be described in \autoref{sec:gnn},
    the same mapping is applied to all \glspl{ap} and users.
    This is a significant advantage in large network scenarios with many \glspl{ap} and users.
    By optimizing the mapping~$N_\theta$ instead of $\mathbf{S}$,
    the complexity becomes independent of the numbers of \glspl{ap} and users.
    This achieves a high scalability,
    which is a major challenge in \gls{cf}.
\end{enumerate}

Formulation~\eqref{eq:ml} presents a general approach.
In the following sections,
we propose specialized \gls{gnn} architecture and
objective function according to the problem properties described in \autoref{sec:properties}.
\section{Hierarchically Permutation-Equivariant Graph Neural Network}
\label{sec:gnn}

\begin{figure}[htbp]
    \centering
    \begin{tikzpicture}[
roundnode/.style={circle, draw=black, minimum size=7mm},
]
\tikzstyle{every node}=[font=\footnotesize]

\node[roundnode] (node1) {AP 1};
\node[roundnode] (node2) [right= 3cm of node1, yshift=0cm] {AP 2};
\node[roundnode] (node3) [above = 0.2 of node1, xshift=1.5cm] {AP 3};

\draw[->] ([yshift=2cm]node2) -- node [above, yshift=0cm] {$\mathbf{e}_{12}$} (node1);
\draw[->] ([yshift=-2cm]node1) -- node [below, yshift=0cm] {$\mathbf{e}_{21}$} (node2);


\draw[->] ([xshift=1cm]node3) -- node [right] {$\mathbf{e}_{31}$} (node1);
\draw[->] ([xshift=-1cm]node1) -- node [left] {$\mathbf{e}_{13}$} (node3);





\draw[->] ([yshift=1cm]node2) -- node [above] {$\mathbf{e}_{23}$} (node3);
\draw[->] ([yshift=-1cm]node3) -- node [below, xshift=-.3cm, yshift=.2cm] {$\mathbf{e}_{32}$} (node2);

\end{tikzpicture}
    \caption{An illustration of a graph.
    Each node in the graph represents an \gls{ap} and edges between nodes indicate cooperation between \glspl{ap}.}
    \label{fig:graph}
\end{figure}
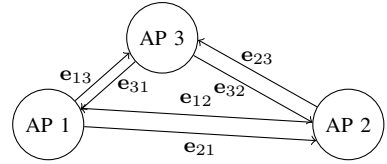

The \gls{gnn} structure is applied for \gls{cf} optimization due to its scalability and decentralized nature.
A \gls{gnn} performs inference on a graph
consisting of nodes and edges,
as illustrated in \autoref{fig:graph}.
In our scenario, each node in the graph represents an \gls{ap}, and the edges between nodes indicate cooperation between \glspl{ap}.

In layer~$i$ of the \gls{gnn},
inference at node~$n$ is performed as
\begin{equation}
    \mathbf{F}_n^{(i+1)}=\gamma^{(i)}\left( \mathbf{F}_n^{(i)}, \oplus_{m\in \mathcal{N}(n)}\phi^{(i)}\left( \mathbf{F}_m^{(i)}, \mathbf{e}_{mn} \right) \right),
    \label{eq:inference}
\end{equation}
where 
$\mathbf{F}_n^{(i+1)}$ is the node feature of node~$n$ as the output of layer~$i$,
$\gamma^{(i)}$ is an \gls{nn} as node processing unit of layer~$i$,
$\mathbf{F}_n^{(i)}$ is the node feature of node~$n$ as the input of layer~$i$,
$\oplus_{m\in \mathcal{N}(n)}$ is an aggregation function over all neighboring nodes of node~$n$ 
(in this work, averaging is used),
$\phi^{(i)}$ is an \gls{nn} as edge processing unit of layer~$i$,
$\mathbf{e}_{mn}$ is the constant edge feature of the edge from node~$m$ to node~$n$.
Inference~\eqref{eq:inference} can be interpreted as three steps:
In Step~1, node~$m$ uses $\phi^{(i)}$ to generate a message 
for its neighboring node~$n$ using its node feature~$\mathbf{F}_m^{(i)}$ and edge feature~$\mathbf{e}_{mn}$.
In Step~2, all messages from neighbors of node~$n$ are aggregated using operator~$\oplus$.
In Step~3, 
node~$n$ uses $\gamma^{(i)}$ to generate its node feature~$\mathbf{F}_n^{(i+1)}$ as output of layer~$i$ according to $\mathbf{F}_n^{(i)}$ and the aggregated message computed in Step~2.

Compared to the generic \gls{gnn} model~\cite{fey1903fast},
a key difference in our model~\eqref{eq:inference} is that $\phi^{(i)}$ does not take $\mathbf{F}_n^{(i)}$ as an argument,
such that the message from node~$m$ to node~$n$ is computed with information locally available at node~$m$.
Therefore,
data traffic in fronthaul between \glspl{ap} is significantly less than in the generic \gls{gnn} model.

Note that $\gamma^{(i)}$ and $\phi^{(i)}$ are applied to all nodes and edges, respectively.
Therefore,
the number of trainable \gls{gnn} parameters is independent of the number of nodes and edges, resulting in good scalability.

Although \eqref{eq:inference} ensures permutation-equivariance at the \gls{ap} level, it does not achieve permutation-equivariance at the user level when using a conventional \gls{nn} architecture, such as a fully connected \gls{nn}, for $\phi$ and $\gamma$~\cite{fey1903fast}.
In this work, we employ a dedicated architecture similar to the RISnet in~\cite{peng2023risnet2},
which is permutation-equivariant to users.
Its number of trainable parameters is also independent of the number of users.

\begin{figure}[htbp]
    \centering
    \subfigure[First layer\label{fig:sinrnet1}]{\resizebox{.92\linewidth}{!}{\begin{tikzpicture}
\tikzstyle{layer} = [rectangle, rounded corners, minimum width=2.5cm, minimum height=.6cm, align=center, text centered, draw=black]

\draw[thick, yshift=.5cm] (0, 0) grid +(4, 1);

\node[font=\LARGE] at (2, -.5) {Users};
\node[font=\LARGE] at (-1.5, 1) {Input};

\draw [decorate,decoration={brace,amplitude=5pt,mirror}]
(-3,-1) -- (4.25,-1) node[midway, yshift=-.7cm, font=\LARGE]{Input matrix};
\draw [decorate,decoration={brace,amplitude=5pt,mirror}]
(4.75,-1) -- (7.5,-1) node[midway, yshift=-.7cm, font=\LARGE]{Info. proc.};
\draw [decorate,decoration={brace,amplitude=5pt,mirror}]
(8,-1) -- (15.1,-1) node[midway, yshift=-.7cm, font=\LARGE]{Output matrix};

\node[draw,thick,rounded corners,rotate=90,inner sep=0em,minimum width=2cm,minimum height=1.5cm] (filters) at (6., 1) {};
\node (layer1) [xshift=6cm, yshift=1.5cm,font=\LARGE] {\texttt{c}};
\node (layer2) [below of=layer1, yshift=0cm,font=\LARGE] {\texttt{a}};
\draw [->,thick] (layer1.east) -- (7.75, 1.5);
\draw [->,thick] (layer2.east) -- (7.75, 0.5);
\draw [->,thick] (4.25, 1) -- ++ (.95, 0);

\draw[thick] (11, 0) grid +(4, 2);
\node[font=\LARGE] at (13, -.5) {Users};
\node[font=\LARGE] at (9.5, 1.5) {Feature \texttt{c}};
\node[font=\LARGE] at (9.5, .5) {Feature \texttt{a}};
\end{tikzpicture}}}
    
    \subfigure[Second layer\label{fig:sinrnet2}]{\resizebox{.92\linewidth}{!}{\begin{tikzpicture}
\tikzstyle{layer} = [rectangle, rounded corners, minimum width=2.5cm, minimum height=.6cm, align=center, text centered, draw=black]

\draw[thick] (0, 0) grid +(4, 2);

\node[font=\LARGE] at (2, -.5) {Users};
\node[font=\LARGE] at (-1.5, 1.5) {Feature \texttt{c}};
\node[font=\LARGE] at (-1.5, .5) {Feature \texttt{a}};

\draw [decorate,decoration={brace,amplitude=5pt,mirror}]
(-3,-1) -- (4.25,-1) node[midway, yshift=-.7cm, font=\LARGE]{Input matrix};
\draw [decorate,decoration={brace,amplitude=5pt,mirror}]
(4.75,-1) -- (7.5,-1) node[midway, yshift=-.7cm, font=\LARGE]{Info. proc.};
\draw [decorate,decoration={brace,amplitude=5pt,mirror}]
(8,-1) -- (15.1,-1) node[midway, yshift=-.7cm, font=\LARGE]{Output matrix};

\node[draw,thick,rounded corners,rotate=90,inner sep=0em,minimum width=2cm,minimum height=2cm] (filters) at (6.25, 1) {};
\node (layer1) [xshift=6cm, yshift=1.0cm,font=\LARGE, text width=1cm] {Info. proc.};
\draw [->,thick] (filters.south) -- (8.5, 1.0);
\draw [->,thick] (4.2, 1) -- ++ (.95, 0);

\draw[thick, yshift=.5cm] (11, 0) grid +(4, 1);
\node[font=\LARGE] at (13, -.5) {Users};
\node[font=\LARGE] at (9.5, 1.0) {Output};
\end{tikzpicture}}}
    \caption{Information processing (info. proc.) of $\gamma$ and $\phi$.}
    \label{fig:info_proc}
\end{figure}
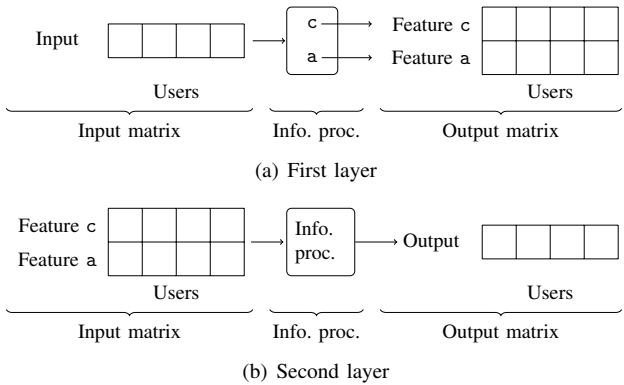

Both $\gamma$ and $\phi$ in all layers of the \gls{gnn} (therefore without superscripts) have two layers.
The inputs of $\gamma$ and $\phi$ are two dimensional  matrices,
where the first dimension is the feature and
the second dimension is the user,
i.e., every user has a feature vector as a column of the input matrix.
Since the decision whether to serve a user depends on both this user (the \emph{current user})
and other users,
we define an information processing unit~\texttt{c} for the \texttt{c}urrent user
and an information processing unit~\texttt{a} for \texttt{a}ll users.
The output feature of Layer~1 for user~$k$ is computed as
\begin{equation}
\mathbf{f}_{k, 2} =
\begin{pmatrix}
     \text{ReLU}(\mathbf{W}^{\texttt{c}}_{1} \mathbf{f}_{k, 1} + \mathbf{b}_1^{\texttt{c}}) \\
     \big(\sum_{k'}\text{ReLU}(\mathbf{W}^{\texttt{a}}_{1} \mathbf{f}_{k', 1} + \mathbf{b}_1^{\texttt{a}})\big) \big/ K\\
\end{pmatrix},
\label{eq:layer_processing1}
\end{equation}
where $\mathbf{W}_1^{\texttt{c}}$ and $\mathbf{b}_1^{\texttt{c}}$ are weights and bias of information processing unit~\texttt{c} on Layer~1, 
respectively,
and $\mathbf{W}_1^{\texttt{a}}$ and $\mathbf{b}_1^{\texttt{a}}$ are weights and bias of information processing unit~\texttt{a} on Layer~1,
respectively.
It can be seen that the output for user~$k$ using information processing unit~\texttt{c} depends only on the input feature of user~$k$,
whereas the output for user~$k$ using information processing unit~\texttt{a} is averaged over all users.
In the second layer,
we do not use two information processing units to reduce data traffic in fronthaul. 
The output feature of user~$k$ is computed as
\begin{equation}
    \mathbf{f}_{k,3}=\text{ReLU}(\mathbf{W}_2 \mathbf{f}_{k,2} + \mathbf{b}_2),
\label{eq:layer_processing2}
\end{equation}
where $\mathbf{W}_2$, $\mathbf{b}_2$ are weights and bias of Layer~2, respectively.
The above-described information processing of $\gamma$ and $\phi$ is illustrated in \autoref{fig:info_proc}.

We let every \gls{ap} make $U$~decisions (serving at most $U$~users according to \eqref{eq:upper}).
Following the principle of \glspl{rnn},
we run the \gls{gnn} $U$~times.
At the $u$-th run,
a relaxed decision of assignment~$s_{kn}^{(u)} \in [0, 1]$ between \gls{ap}~$n$ and user~$k$ is made.
The activation function softmax is applied in the final layer such that $s_{kn}^{(u)} \in [0, 1]$ for all $n$, $k$ and $u$,
and $\sum_{k=1}^K s_{kn}^{(u)} = 1$ for all $n$ and $u$.
The input of user~$k$ and \gls{ap}~$n$ at the $u$-th run comprises 
the channel gain~$g_{kn}$
and gap to fulfill constraint \eqref{eq:lower} after $u$~runs,
i.e., $\text{ReLU}(L - \sum_{\mu=1}^u s_{kn}^{(\mu)})$.

\section{Training with Augmented Lagrangian Method}
\label{sec:training}

In this section,
we propose a training process for Problem~\eqref{eq:problem} using \gls{alm}.
Unlike conventional applications of \gls{alm},
which tune optimization variables directly,
we use \gls{alm} to formulate the objective,
then use the Adam optimizer to optimize the \gls{nn} parameter~$\theta$.
During training,
$s_{kn}$ is relaxed to a continuous variable between 0 and 1.
Constraint~\eqref{eq:upper} is guaranteed by the recurrent process described in \autoref{sec:gnn}.
Constraints~\eqref{eq:discrete} and \eqref{eq:lower} 
appear in the objective as penalty terms.
Using \gls{alm},
the objective function is defined as
\begin{equation}
\begin{split}
g &= f - \lambda_1 \sum_{k=1}^K \text{ReLU}\left(L - \sum_{n=1}^N s_{kn}\right) - \frac{1}{2}\lambda_2 \sum_{n=1}^N p_n\\
&\qquad -\nu_1 \sum_{k=1}^K \text{ReLU} \left(L - \sum_{n=1}^N s_{kn}\right)^2 - \frac{1}{2}\nu_2 \sum_{n=1}^N p_n^2,
\end{split}
\label{eq:alm}
\end{equation}
where 
$\lambda_j$ and $\nu_j$ are the factors of the linear and
quadratic penalty terms,
respectively ($j=1, 2$),
$p_n$ is the penalty term of \gls{ap}~$n$ for \eqref{eq:discrete} inspired by the information entropy:
\begin{equation}
    p_n = -\sum_{k=1}^K s_{kn} \log(s_{kn}).
    \label{eq:entropy}
\end{equation}
Note that $p_n$ is equal to 0 if $s_{kn}=0$ or $1$.
Otherwise $p_n > 0$.

The training is described in \autoref{alg},
where $\Delta \nu$ is a predefined constant changing rate of $\nu_1$ and $\nu_2$.

\begin{algorithm}[htbp]
\caption{Training using \gls{alm}}
\label{alg}
\begin{algorithmic}[1]
\State Set $\lambda_j = \nu_j = 0$ for $j=1, 2$.
\Repeat
\State Update \gls{gnn} with a gradient ascent step.
\Until{Convergence}

\Repeat
\State $\lambda_1 \leftarrow \lambda_1 + \nu_1 \sum_{k=1}^K \text{ReLU}\left(L - \sum_{n=1}^N s_{kn}\right)$
\State $\nu_1 \leftarrow \nu_1 + \Delta\nu$
\Repeat
\State Update \gls{gnn} with a gradient ascent step.
\Until{Convergence}
\Until{$\sum_{k=1}^K \text{ReLU}\left(L - \sum_{n=1}^N s_{kn}\right) = 0$.}

\Repeat
\State $\lambda_2 \leftarrow \lambda_2 + \nu_2 \sum_{k=1}^K \text{ReLU}\left(L - \sum_{n=1}^N s_{kn}\right)$
\State $\nu_2 \leftarrow \nu_2 + \Delta\nu$
\Repeat
\State Update \gls{gnn} with a gradient ascent step.
\Until{Convergence}
\Until{$-\sum_{k=1}^Ks_{kn}\log(s_{kn})=0$.}
\end{algorithmic}
\end{algorithm}

Note that we added constraints \eqref{eq:lower} and \eqref{eq:entropy} one after another.
This is different from canonical \gls{alm},
which updates $\lambda_j$ and $\nu_j$ simultaneously for all constraints.
However, training results show that the proposed method guarantees the constraints better.
\section{Training and Testing Results}
\label{sec:results}

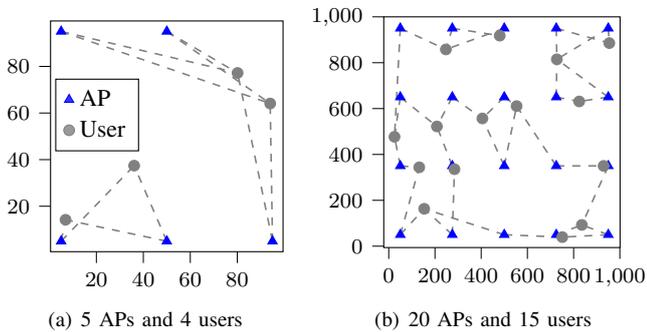
\begin{figure}[htbp]
    \centering
    \subfigure[5 \glspl{ap} and 4 users]{%
\begin{tikzpicture}

\definecolor{darkgray176}{RGB}{176,176,176}
\definecolor{gray}{RGB}{128,128,128}
\definecolor{green}{RGB}{128,128,128}
\definecolor{lightgray204}{RGB}{204,204,204}

\begin{axis}[
	scale only axis,
	width=.4\linewidth,
	ticklabel style={font=\footnotesize},
	axis equal image,
legend cell align={left},
legend style={
  fill opacity=0.8,
  draw opacity=1,
  text opacity=1,
  at={(.02,.75)},
  anchor=north west,
  draw=black,
  font=\small,
},
tick align=outside,
tick pos=left,
x grid style={darkgray176},
xmin=0.5, xmax=99.5,
xtick style={color=black},
y grid style={darkgray176},
ymin=0.5, ymax=99.5,
ytick style={color=black},
]
\addplot [draw=blue, fill=blue, mark=triangle*, only marks]
table{%
x  y
5 5
50 5
95 5
5 95
50 95
};
\addlegendentry{AP}

\addplot [draw=gray, fill=gray, mark=*, only marks]
table{%
x  y
6.8810133934021 14.1929883956909
94.0179214477539 64.0831146240234
80.1267929077148 77.1919555664062
36.0946006774902 37.4101905822754
};
\addlegendentry{User}

\addplot [semithick, dashed, green, forget plot]
table {%
5 5
6.8810133934021 14.1929883956909
};
\addplot [semithick, dashed, green, forget plot]
table {%
50 5
6.8810133934021 14.1929883956909
};
\addplot [semithick, dashed, green, forget plot]
table {%
95 5
80.1267929077148 77.1919555664062
};
\addplot [semithick, dashed, green, forget plot]
table {%
5 95
94.0179214477539 64.0831146240234
};
\addplot [semithick, dashed, green, forget plot]
table {%
50 95
94.0179214477539 64.0831146240234
};
\addplot [semithick, dashed, green, forget plot]
table {%
5 5
36.0946006774902 37.4101905822754
};
\addplot [semithick, dashed, green, forget plot]
table {%
50 5
36.0946006774902 37.4101905822754
};
\addplot [semithick, dashed, green, forget plot]
table {%
95 5
94.0179214477539 64.0831146240234
};
\addplot [semithick, dashed, green, forget plot]
table {%
5 95
80.1267929077148 77.1919555664062
};
\addplot [semithick, dashed, green, forget plot]
table {%
50 95
80.1267929077148 77.1919555664062
};
\end{axis}

\end{tikzpicture}}
    \subfigure[20 \glspl{ap} and 15 users]{%
\begin{tikzpicture}

\definecolor{darkgray176}{RGB}{176,176,176}
\definecolor{gray}{RGB}{128,128,128}
\definecolor{green}{RGB}{0,128,0}
\definecolor{lightgray204}{RGB}{204,204,204}

\begin{axis}[
scale only axis,
width=.4\linewidth,
ticklabel style={font=\footnotesize},
axis equal image,
legend cell align={left},
legend style={
	fill opacity=0.8,
	draw opacity=1,
	text opacity=1,
	at={(1.01,1)},
	anchor=north west,
	draw=black
},
tick align=outside,
tick pos=left,
x grid style={darkgray176},
xmin=-21.5757002830505, xmax=1000.89347410202,
xtick style={color=black},
y grid style={darkgray176},
ymin=-5.98994445800781, ymax=1000,
ytick style={color=black}
]
\addplot [draw=blue, fill=blue, mark=triangle*, only marks]
table{%
	x  y
	50 50
	275 50
	500 50
	725 50
	950 50
	50 350
	275 350
	500 350
	725 350
	950 350
	50 650
	275 650
	500 650
	725 650
	950 650
	50 950
	275 950
	500 950
	725 950
	950 950
};

\addplot [draw=gray, fill=gray, mark=*, only marks]
table{%
	x  y
	132.008285522461 344.017517089844
	24.9001712799072 476.909881591797
	153.391479492188 163.114044189453
	835.628662109375 92.7466506958008
	208.399810791016 522.105895996094
	824.204467773438 631.187133789062
	405.296966552734 557.046447753906
	553.254150390625 610.939025878906
	284.386474609375 335.652435302734
	929.8125 350.140014648438
	954.417602539062 885.551940917969
	247.25439453125 857.893798828125
	750.965026855469 39.5333862304688
	480.32666015625 917.894409179688
	727.464111328125 814.809265136719
};

\addplot [semithick, dashed, gray, forget plot]
table {%
	50 50
	132.008285522461 344.017517089844
};
\addplot [semithick, dashed, gray, forget plot]
table {%
	275 50
	153.391479492188 163.114044189453
};
\addplot [semithick, dashed, gray, forget plot]
table {%
	500 50
	750.965026855469 39.5333862304688
};
\addplot [semithick, dashed, gray, forget plot]
table {%
	725 50
	750.965026855469 39.5333862304688
};
\addplot [semithick, dashed, gray, forget plot]
table {%
	950 50
	750.965026855469 39.5333862304688
};
\addplot [semithick, dashed, gray, forget plot]
table {%
	50 350
	132.008285522461 344.017517089844
};
\addplot [semithick, dashed, gray, forget plot]
table {%
	275 350
	284.386474609375 335.652435302734
};
\addplot [semithick, dashed, gray, forget plot]
table {%
	500 350
	553.254150390625 610.939025878906
};
\addplot [semithick, dashed, gray, forget plot]
table {%
	725 350
	553.254150390625 610.939025878906
};
\addplot [semithick, dashed, gray, forget plot]
table {%
	950 350
	929.8125 350.140014648438
};
\addplot [semithick, dashed, gray, forget plot]
table {%
	50 650
	24.9001712799072 476.909881591797
};
\addplot [semithick, dashed, gray, forget plot]
table {%
	275 650
	208.399810791016 522.105895996094
};
\addplot [semithick, dashed, gray, forget plot]
table {%
	500 650
	553.254150390625 610.939025878906
};
\addplot [semithick, dashed, gray, forget plot]
table {%
	725 650
	727.464111328125 814.809265136719
};
\addplot [semithick, dashed, gray, forget plot]
table {%
	950 650
	727.464111328125 814.809265136719
};
\addplot [semithick, dashed, gray, forget plot]
table {%
	50 950
	24.9001712799072 476.909881591797
};
\addplot [semithick, dashed, gray, forget plot]
table {%
	275 950
	247.25439453125 857.893798828125
};
\addplot [semithick, dashed, gray, forget plot]
table {%
	500 950
	480.32666015625 917.894409179688
};
\addplot [semithick, dashed, gray, forget plot]
table {%
	725 950
	727.464111328125 814.809265136719
};
\addplot [semithick, dashed, gray, forget plot]
table {%
	950 950
	727.464111328125 814.809265136719
};
\addplot [semithick, dashed, gray, forget plot]
table {%
	50 50
	153.391479492188 163.114044189453
};
\addplot [semithick, dashed, gray, forget plot]
table {%
	275 50
	284.386474609375 335.652435302734
};
\addplot [semithick, dashed, gray, forget plot]
table {%
	500 50
	153.391479492188 163.114044189453
};
\addplot [semithick, dashed, gray, forget plot]
table {%
	725 50
	835.628662109375 92.7466506958008
};
\addplot [semithick, dashed, gray, forget plot]
table {%
	950 50
	835.628662109375 92.7466506958008
};
\addplot [semithick, dashed, gray, forget plot]
table {%
	50 350
	24.9001712799072 476.909881591797
};
\addplot [semithick, dashed, gray, forget plot]
table {%
	275 350
	208.399810791016 522.105895996094
};
\addplot [semithick, dashed, gray, forget plot]
table {%
	500 350
	405.296966552734 557.046447753906
};
\addplot [semithick, dashed, gray, forget plot]
table {%
	725 350
	929.8125 350.140014648438
};
\addplot [semithick, dashed, gray, forget plot]
table {%
	950 350
	835.628662109375 92.7466506958008
};
\addplot [semithick, dashed, gray, forget plot]
table {%
	50 650
	208.399810791016 522.105895996094
};
\addplot [semithick, dashed, gray, forget plot]
table {%
	275 650
	405.296966552734 557.046447753906
};
\addplot [semithick, dashed, gray, forget plot]
table {%
	500 650
	405.296966552734 557.046447753906
};
\addplot [semithick, dashed, gray, forget plot]
table {%
	725 650
	824.204467773438 631.187133789062
};
\addplot [semithick, dashed, gray, forget plot]
table {%
	950 650
	824.204467773438 631.187133789062
};
\addplot [semithick, dashed, gray, forget plot]
table {%
	50 950
	247.25439453125 857.893798828125
};
\addplot [semithick, dashed, gray, forget plot]
table {%
	275 950
	480.32666015625 917.894409179688
};
\addplot [semithick, dashed, gray, forget plot]
table {%
	500 950
	247.25439453125 857.893798828125
};
\addplot [semithick, dashed, gray, forget plot]
table {%
	725 950
	954.417602539062 885.551940917969
};
\addplot [semithick, dashed, gray, forget plot]
table {%
	950 950
	954.417602539062 885.551940917969
};
\end{axis}
\end{tikzpicture}}
    \caption{Visualization of \gls{ap}-user assignment in two scenarios.}
    \label{fig:visualization}
\end{figure}

For evaluating the training and testing performance, we define a small and a large network scenario,
where
\glspl{ap} are uniformly placed (deterministically)
and users are randomly distributed,
as shown in \autoref{fig:visualization}.
The user positions in the training and testing sets are generated independently but with the same uniform distributions.
The path loss is assumed to be Rician distributed,
where the expectation is reciprocal to the distance from \gls{ap} to user,
and the variance is constant.
Important scenario parameters are listed in \autoref{tab:parameters}.

\begin{table}[htbp]
    \centering
    \caption{Important scenario parameters.}
    \label{tab:parameters}
    \begin{tabular}{p{.50\linewidth}p{.40\linewidth}}
    \toprule
    Parameter & Value\\
    \midrule
        Number of \glspl{ap} & 20 (large network scenario),\newline{}5 (small network scenario) \\
        Number of users & 15 (large network scenario),\newline{}4 (small network scenario) \\
        Minimum number of serving \glspl{ap} per user~$L$ & 2 \\
        Maximum number of served users per \gls{ap}~$U$ & 2\\
    \bottomrule
    \end{tabular}
\end{table}

\begin{figure}
    \centering
    \input{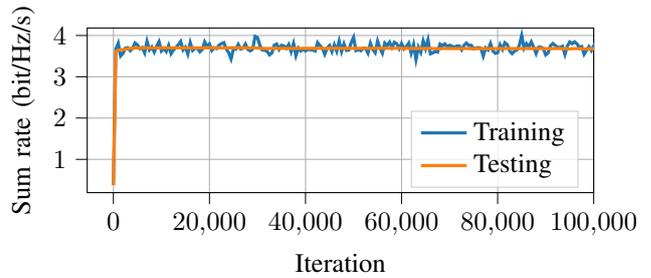}
    \vspace*{-1em} 
    \caption{Sum-rate in training and testing over iterations.}
    \label{fig:sum-rate}
\end{figure}

We perform testing on a testing set (which is disjoint from the training set) after every iteration of training.
\autoref{fig:sum-rate} shows realized curves of the sum rate for both training and testing phases.
The sum rate in the testing phase improves quickly to \SI{3.70}{\bit\per\Hz\per\s} without constraints, and
then decreases slightly to \SI{3.67}{\bit\per\Hz\per\s} due to the constraints.
The training curve fluctuates while the testing curve remains smooth due to the fact that a random batch is sampled from the training set in each iteration, whereas the same testing set is used for every testing iteration.
Despite the fluctuation in the training curve,
the good match between the training and testing curves on average indicates that there is no overfitting.
This is because the same information processing units are applied to all \gls{ap}-user connections
(see \autoref{sec:gnn}).
Therefore, the \gls{gnn} is relatively small compared to the problem size.

\begin{figure}
    \centering
    \input{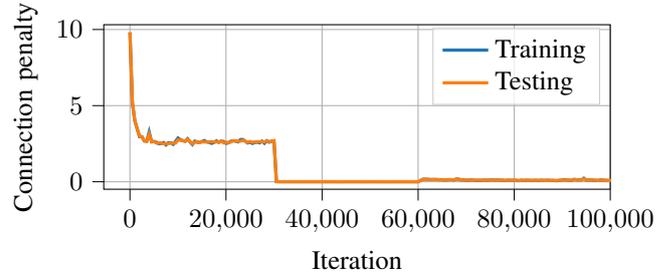}
    \vspace*{-2em}
    \caption{Penalty term of constraint \eqref{eq:lower} in training and testing over iterations.}
    \label{fig:conn}
\end{figure}

\autoref{fig:conn} shows the realized connection penalty of users \eqref{eq:lower} during both training and testing.
After introducing the penalty for constraint~\eqref{eq:lower} (i.e., Line~5 in \autoref{alg}),
the connection penalty is reduced to 0.
After introducing the penalty for discreteness~\eqref{eq:entropy} 
(i.e., Line~11 in \autoref{alg}),
the connection penalty slightly increases before slowly decreasing to 0.
We confirm that constraint~\eqref{eq:lower} is satisfied in all 1024 testing samples at the end of training.

\begin{figure}
    \centering
    \input{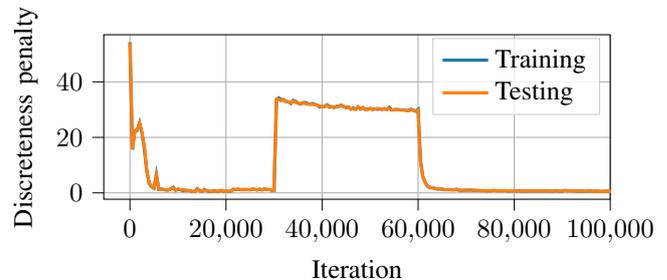}
    \vspace*{-2em}
    \caption{Discreteness penalty in training and testing over iterations.}
    \label{fig:discreteness}
\end{figure}

The training and testing curves of the discreteness penalty~\eqref{eq:entropy} are shown in \autoref{fig:discreteness}.
After introducing the penalty for discreteness
(i.e., Line~11 in \autoref{alg}),
the discreteness penalty reduces to 0.
Therefore, the solution to the relaxed problem is valid for the original problem~\eqref{eq:problem}.

Two visualizations of small and large network scenarios are shown in \autoref{fig:visualization},
with dashed lines indicating the \gls{ap}-user assignment.
The assignments match our intuition
because users are always served by nearby \glspl{ap}, subject to constraints~\eqref{eq:upper} and \eqref{eq:lower}.

\autoref{tab:comparison} presents a comparison of sum spectral efficiency obtained with the proposed method, an upper bound (found via exhaustive search), 
the average performance of 100~random assignments,
and the \gls{gsd} algorithm~\cite{cechlarova2014pareto}.
When performing an exhaustive search,
we evaluate $\binom{K}{U}^N$ possible assignments while considering constraints~\eqref{eq:discrete} and \eqref{eq:upper}.
This results in \num{7776} and $2.6\times 10^{40}$ assignments in the small and large network scenarios, respectively.
The proposed method achieves a performance very close to the upper bound in the small network scenario, whereas the upper bound cannot be calculated in the large network scenario due to limited memory and time.
In both scenarios,
the proposed method outperforms the random assignment and \gls{gsd} algorithm.
Additionally,
the proposed method provides a solution within a few milliseconds
with reduced data traffic in fronthaul
(see \autoref{sec:gnn}),
and without a centralized processing unit,
indicating potential for real-time application.

\begin{table}[htbp]
    \centering
    \caption{Achieved sum spectral efficiencies (bit/Hz/s) of proposed method and baselines.}
    \label{tab:comparison}
    \begin{tabular}{lll}
    \toprule
      & Small scenario & Large scenario \\
    \midrule
    Proposed & 1.14 & 3.67\\
    Upper bound & 1.15 & Not available \\
    Random & 0.60 & 0.51 \\
    GSD & 1.06 & 3.34\\
    \bottomrule
    \end{tabular}
\end{table}
\section{Conclusion}
\label{sec:conclusion}

Due to challenging \gls{mmwave} channel and hardware characteristics,
the \gls{mmwave} \gls{cf} optimization problem is primarily an assignment problem between \glspl{ap} and users.
In this work,
we have proposed an unsupervised \gls{ml} approach.
In particular, we have utilized a customized \gls{gnn} to achieve scalability, distributed inference, hierarchical permutation-equivariance,
and reduced data traffic in fronthaul.
To solve the combinatorial assignment problem,
we first relaxed it to a continuous problem.
Next, we introduced a penalty inspired by information entropy and use \gls{alm} to enforce discreteness.
Testing results indicate that the proposed method
outperforms both random assignment and \gls{gsd} algorithm.
In a small network scenario, its performance is very close to the upper bound found via exhaustive search.
In a large network scenario, where exhaustive search is not feasible due to memory and time limitations, the proposed method still achieves good performance.
The proposed method could be generalized to other large-scale combinatorial problems in the future.
The combination of \gls{ml} and analytical \gls{alm} can also be applied to other constrained optimization problems.

Code and data of this work will be made public on Github if the paper is accepted.

\printbibliography

\end{document}